\documentclass[leqno]{amsart}

\usepackage[foot]{amsaddr}

\usepackage{color}
\usepackage{geometry}
\usepackage{subfiles}
\usepackage{graphicx}
\usepackage[section]{placeins}
\usepackage{hyperref}
\usepackage{stmaryrd}
\usepackage{amsmath,amssymb,amsfonts,amsthm,textcomp}
\usepackage{mathtools}
\usepackage{tensor}
\usepackage{multicol}
\usepackage{subfig}
\graphicspath{
              {./figures_section-1/}
              {./figures_section-2/}
              {./figures_section-3/}
             }

\newfont{\tenbfsl}{cmbxti9 scaled 1200}
\newfont{\tenbbb}{msbm10}
\newfont{\svnbbb}{msbm8}

\newcommand{\bs}[1]{\boldsymbol{#1}}
\newcommand{\br}[1]{\boldsymbol{\mathrm{{#1}}}}

\newcommand{\cl}[1]{\mathcal{#1}}

\newcommand{\bb}[1]{\mathbb{#1}}

\newcommand{\id}{\bs{1}}

\newcommand{\surp}[1]{\{\!\!\{ {#1} \}\!\!\}}

\newcommand{\fr}[2]{{\textstyle{\frac{{#1}}{{#2}}}}}

\newcommand{\dv}{\,\mathrm{d}v}
\newcommand{\da}{\,\mathrm{d}a}
\newcommand{\ds}{\,\mathrm{d}\sigma}

\newcommand{\bdy}{\cl{B}}
\newcommand{\dbdy}{\partial\cl{B}}
\newcommand{\prt}{\cl{P}}
\newcommand{\dprt}{\partial\cl{P}}
\newcommand{\ddprt}{\partial^2\cl{P}}
\newcommand{\srf}{\cl{S}}
\newcommand{\dsrf}{\partial\cl{S}}

\newcommand{\edg}{\cl{C}}

\newcommand{\intP}{\int\limits_{\prt}}
\newcommand{\intdP}{\int\limits_{\dprt}}
\newcommand{\intddP}{\int\limits_{\ddprt}}
\newcommand{\intoddP}{\int\limits_{{}_\circ\ddprt}}

\newcommand{\trans}{\scriptscriptstyle\mskip-1mu\top\mskip-2mu}

\newcommand{\Grad}{\mathrm{grad}\mskip2mu}
\newcommand{\Div}{\mathrm{div}\mskip2mu}

\newcommand{\Grads}{\Grad_{\mskip-2mu\scriptscriptstyle\cl{S}}}
\newcommand{\Divs}{\Div_{\mskip-6mu\scriptscriptstyle\cl{S}}}
\newcommand{\triangles}{\triangle_{\mskip-2mu\scriptscriptstyle\cl{S}}}

\theoremstyle{remark}

\theoremstyle{definition}

\newcommand{\twovdots}{\mskip+2mu\colon\mskip-2mu}
\makeatletter
\def\threevdots{\mskip+4mu\vbox{\baselineskip2.25\p@ \lineskiplimit\z@
  \kern4.9\p@\hbox{.}\hbox{.}\hbox{.}}\mskip+3.8mu}
\makeatother

\newcommand{\Prj}[1]{\br{P}_{\scriptscriptstyle{\mskip-6mu{#1}}}}

\newcommand{\xis}{\xi_{\scriptscriptstyle\cl{S}}}
\newcommand{\tauds}{\tau_{\scriptscriptstyle\partial\cl{S}}}
\newcommand{\otauds}{{}_\circ\!\tau_{\scriptscriptstyle\partial\cl{S}}}

\newcommand{\chiP}{\chi_{\scriptscriptstyle\cl{P}}}
\newcommand{\chidP}{\chi_{\scriptscriptstyle\partial\cl{P}}}

\newcommand{\psiP}{\psi_{\scriptscriptstyle\cl{P}}}
\newcommand{\psidP}{\psi_{\scriptscriptstyle\partial\cl{P}}}
\newcommand{\dpsiP}{\dot{\psi}_{\scriptscriptstyle\cl{P}}}
\newcommand{\dpsidP}{\dot{\psi}_{\scriptscriptstyle\partial\cl{P}}}

\newcommand{\vphiP}{\varphi_{\scriptscriptstyle\cl{P}}}
\newcommand{\vphidP}{\varphi_{\scriptscriptstyle\partial\cl{P}}}
\newcommand{\vphiddP}{\varphi_{\scriptscriptstyle\partial^2\cl{P}}}
\newcommand{\dvphiP}{\dot{\varphi}_{\scriptscriptstyle\cl{P}}}
\newcommand{\dvphidP}{\dot{\varphi}_{\scriptscriptstyle\partial\cl{P}}}
\newcommand{\dvphiddP}{\dot{\varphi}_{\scriptscriptstyle\partial^2\cl{P}}}

\newcommand{\bsMP}{\bs{M}_{\mskip-4mu\scriptscriptstyle\cl{P}}}
\newcommand{\bsMdP}{\bs{M}_{\mskip-4mu\scriptscriptstyle\partial\cl{P}}}
\newcommand{\mP}{m_{\mskip-2mu\scriptscriptstyle\cl{P}}}
\newcommand{\mdP}{m_{\mskip-2mu\scriptscriptstyle\partial\cl{P}}}

\newcommand{\sP}{s_{\scriptscriptstyle\cl{P}}}
\newcommand{\sdP}{s_{\scriptscriptstyle\partial\cl{P}}}

\newcommand{\muP}{\mu_{\scriptscriptstyle\cl{P}}}
\newcommand{\mudP}{\mu_{\scriptscriptstyle\partial\cl{P}}}
\newcommand{\muddP}{\mu_{\scriptscriptstyle\partial^2\cl{P}}}

\newcommand{\bsjP}{\bs{\jmath}_{\scriptscriptstyle\cl{P}}}
\newcommand{\bsjdP}{\bs{\jmath}_{\scriptscriptstyle\partial\cl{P}}}

\newcommand{\pvphiP}{\partial_{\vphiP}}
\newcommand{\pgvphiP}{\partial_{\Grad\vphiP}}
\newcommand{\pvphidP}{\partial_{\vphidP}}
\newcommand{\pgvphidP}{\partial_{\Grads\vphidP}}

\begin{document}

\title{A continuum framework for phase field with bulk-surface dynamics}
\author{Luis Espath}
\address{School of Mathematical Sciences, University of Nottingham, Nottingham, NG7 2RD, United Kingdom}
\email{luis.espath@nottingham.ac.uk}

\date{\today}

\begin{abstract}
\noindent
This continuum mechanical theory aims at detailing the underlying rational mechanics of dynamic boundary conditions proposed by Fischer, Maass, \& Dieterich \cite{Fis97}, Goldstein, Miranville, \& Schimperna \cite{Gol11}, and Knopf, Lam, Liu \& Metzger, \cite{Kno21a}. As a byproduct, we generalize these theories. These types of dynamic boundary conditions are described by the coupling between the bulk and surface partial differential equations for phase fields. Our point of departure within this continuum framework is the principle of virtual powers postulated on an arbitrary part $\prt$ where the boundary $\dprt$ may lose smoothness. That is, the normal field may be discontinuous along an edge $\ddprt$. However, the edges characterizing the discontinuity of the normal field are considered smooth. Our results may be summarized as follows. We provide a generalized version of the principle of virtual powers for the bulk-surface coupling along with a generalized version of the partwise free-energy imbalance. Next, we derive the explicit form of the surface and edge microtractions along with the field equations for the bulk and surface phase fields. The final set of field equations somewhat resembles the Cahn--Hilliard equation for both the bulk and surface. Moreover, we provide a suitable set of constitutive relations and thermodynamically consistent boundary conditions. In \cite{Kno21a}, a mixed (Robin) type of boundary condition for the chemical potentials is proposed for the model in \cite{Fis97,Gol11}. In addition to this boundary condition, we also include this type of mixed boundary condition for the microstructure, that is the phase fields. Lastly, we derive the Lyapunov-decay relations for these mixed type of boundary conditions for both the microstructure and chemical potential.
\\
\textbf{AMS subject classifications:}
$\cdot$
74N20 
$\cdot$
80A22 
$\cdot$
80A17 
$\cdot$
82C26 
$\cdot$
35L65 
$\cdot$

\end{abstract}

\maketitle

\tableofcontents                        


\section{Introduction}

Dynamic boundary conditions for phase segregation are ubiquitous in mathematical biology, geology and industrial processes. Conversely, spontaneous phase segregation of binary mixtures has been modeled by the Cahn--Hilliard equation \cite{Cah61}. As for the underlying mechanics of phase segregation, Fried \& Gurtin and Gurtin \cite{Fri93,Gur96} proposed the original continuum framework for the study of these types of equations, namely the Allen--Cahn/Ginzburg--Landau and Cahn--Hilliard equations. Additionally, Espath, Calo \& Fried \cite{Esp20} and Espath \& Calo \cite{Esp21a} generalized these ideas to encompass second gradient theories, namely the Swift--Hohenberg/Brazovski\v{\i} and phase-field crystal equations.

As for the dynamic boundary conditions, a continuum mechanical theory has not yet been proposed to the best of our knowledge. Nonetheless, to account for the types of interactions in the presence of solid walls in confined systems while focusing on the early stage of the demixing kinetics, Fischer, Maass, \& Dieterich \cite{Fis97} proposed a set of dynamic boundary conditions for flat walls. These dynamic boundary conditions are characterized by an evolution equation on the boundary coupled with the bulk's evolution equation. Many phenomena may fit into this scenario, including polymer mixtures, metallic alloys, and metamorphic rock formation, among other physical and industrial processes.

In this work, we aim at exploring the underlying mechanical principles of the bulk-surface connection for phase-field theories. Our continuum framework is constructed based on the work by Fried \& Gurtin \cite{Fri93,Gur96}, Espath \& Calo \cite{Esp21a}, and Espath \cite{Esp21b,Esp21c} to generalize the models proposed by Fischer, Maass, \& Dieterich \cite{Fis97}, Goldstein, Miranville \& Schimperna \cite{Gol11}, and Knopf, Lam, Liu \& Metzger \cite{Kno21a}. To this end, for the bulk-surface coupling, we provide a generalized version of the principle of virtual powers, which allows us to establish meaningful weak forms, with a generalized version of the partwise free-energy imbalance. Next, we derive the explicit form of the surface and edge microtractions along with the field equations for the bulk and surface phase fields. Additionally, through this version of the partwise free-energy imbalance, we propose the constitutive relations and a set of thermodynamically consistent boundary conditions, including mixed (Robin) boundary conditions for the microstructure (describe by the phase fields) and chemical potentials. Moreover, this continuum framework has two different bulk-surface types of couplings, one through the principle of virtual powers and another through the species balance (and consequently through the free-energy imbalance). Finally, we present the Lyapunov-decay relations for a fairly general setting.

The remainder of this work is organized as follows. In Section \S\ref{sc:principle.virtual.powers}, we provide a generalized version of virtual powers for bulk-surface dynamics, and derive the surface and edge microtractions along with the field equations. In Section \S\ref{sc:conserved.species}, we postulate the species balance of the bulk-surface system. In Section \S\ref{sc:free.energy.imbalance}, we postulate a generalized version of the partwise free-energy imbalance. In Section \S\ref{sc:constitutive.relations}, we provide suitable constitutive relations along with a set of thermodynamically consistent boundary conditions. In Section \S\ref{sc:lyapunov}, we derive the Lyapunov-decay relations for mixed (Robin) boundary conditions for both the microstructure and chemical potential.

\subsection{Synopsis of purely variational models}

The model proposed in \cite{Fis97,Gol11,Kno21a} on a body $\prt$ with boundary $\dprt$ for the underlying free-energy functional
\begin{align}\label{eq:free.energy.functional}
\Psi[\vphiP,\vphidP] &= \intP \psiP \dv + \intdP \psidP \da, \nonumber \\[4pt]
&= \intP \Big( \fr{1}{\epsilon} f(\vphiP) + \fr{\epsilon}{2} |\Grad\vphiP|^2 \Big) \dv + \intdP \Big( \fr{1}{\delta} g(\vphidP) + \fr{\iota \delta}{2} |\Grads\vphidP|^2 \Big) \da,
\end{align}
reads
\begin{equation}\label{eq:original}
\left\{
\begin{aligned}
\dvphiP &= \mP\triangle\muP, && \text{in } \prt, \\[4pt]
\muP &= -\epsilon \triangle\vphiP + \fr{1}{\epsilon} f^\prime (\vphiP), && \text{in } \prt, \\[4pt]
\dvphidP &= \mdP\triangles\mudP - \beta \mP \partial_n\muP, && \text{on } \dprt, \\[4pt]
\mudP &= - \iota \delta \triangles\vphidP + \fr{1}{\delta} g^\prime (\vphidP) + \epsilon \partial_n \vphiP, && \text{on } \dprt, \\[4pt]
\vphiP &= \vphidP, && \text{on } \dprt, \\[4pt]
\partial_n \muP &= \fr{1}{L} (\beta \mudP - \muP), && \text{on } \dprt.
\end{aligned}
\right.
\end{equation}
Here, the superposed dot represents the time derivative and when denoted on top of an integral, it represents the total time derivative $\text{d}/\text{d}t$. $\psiP$ and $\psidP$ represent the bulk and surface free-energy densities, respectively. $\triangle$ and $\triangles$ are the Laplace and Laplace--Beltrami operators, respectively. $\vphiP$ and $\vphidP$ are the bulk and surface conserved phase fields, $\muP$ and $\mudP$ are the bulk and surface chemical potentials, $f$ and $g$ are the bulk and surface potentials, and $\mP$ and $\mdP$ are the bulk and surface mobility coefficients. Lastly, $\epsilon$, $\delta$, $\iota$, $\beta$ and $L$ are real positive constant parameters. It is important to what follows to note that \eqref{eq:original} has the Cahn--Hilliard type of structure for both the bulk and surface.

\subsection{Synopsis of this work}

In continuum mechanics, it is customary to isolate an arbitrary part $\prt$ from a body $\bdy$ to describe the interactions between $\prt$ and adjacent parts of $\bdy$ to establish balance laws. That is to say, the action of $\bdy\setminus\prt$ on $\prt$ is represented through surface tractions and normal fluxes. This is probably the most used concept in structural mechanics. We here abandon this hypothesis and consider that interactions between $\prt$ and adjacent parts of $\bdy$ are described by additional evolution equations on $\dprt$. This ultimately implies that the boundary conditions on $\bdy$ are defined through a partial differential equation on $\dbdy$. We may however limit the dynamic response of the environment to a certain region of $\dbdy$ instead of considering that the entire surrounding environment is dynamic. Note that since balances do not depend on material idealizations, we separate balance equations from constitutive response functions.

\begin{figure}[h]
  \centering
  \includegraphics[width=0.65\textwidth]{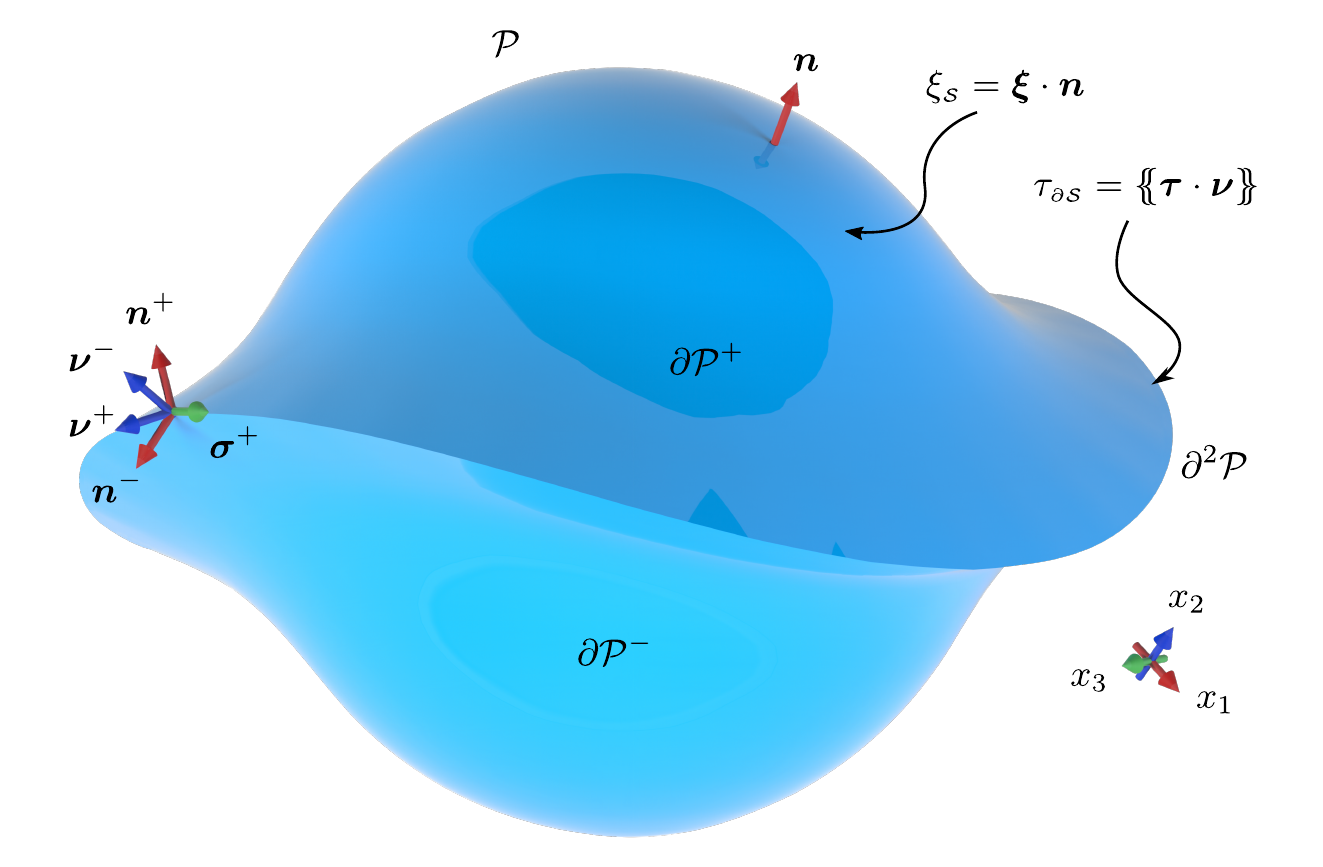}
  \caption{Part $\prt$ with nonsmooth boundary surface $\dprt^\pm$ oriented by the unit normal $\bs{n}$ with the outward unit tangent-normal $\bs{\nu}^\pm$ at the smooth boundary-edge $\ddprt$ oriented by the unit tangent $\bs{\sigma}\coloneqq\bs{n}\times\bs{\nu}$. The surface $\dprt$ lacks smoothness at an edge $\ddprt$.}
  \label{fg:geometry}
\end{figure}

In Figure \ref{fg:geometry}, $\bdy$ denotes a region of a three-dimensional point space $\cl{E}$ where $\prt\subseteq\bdy$ is an arbitrary subregion of $\bdy$ with a closed surface boundary $\dprt$ oriented by an outward unit normal $\bs{n}$ at $\bs{x}\in\dprt$. The surface $\dprt$ may lose smoothness along a curve, namely an edge $\ddprt$. In a neighborhood of an edge $\ddprt$, two smooth surfaces $\dprt^{\pm}$ are defined. The limiting unit normals of $\dprt^{\pm}$ at $\ddprt$ are denoted by the pair $\{\bs{n}^+,\bs{n}^-\}$. The pair of unit normals characterizes the edge $\ddprt$. Similarly, the limiting outward unit tangent-normal\footnote{The unit tangent-normal is a unit vector that is tangent to the surface and normal to the boundary of the surface.} of $\dprt^{\pm}$ at $\ddprt$ are $\{\bs{\nu}^+,\bs{\nu}^-\}$. Additionally, $\ddprt$ is oriented by the unit tangent $\bs{\sigma}\coloneqq\bs{\sigma}^+$ such that $\bs{\sigma}^+\coloneqq\bs{n}^+\times\bs{\nu}^+$. Furthermore, the body $\bdy$ and all its parts are open sets in $\cl{E}$.

In this work, we propose a continuum theory with two kinematical processes, a bulk $\vphiP$ and a surface $\vphidP$ fields on $\prt$ and $\dprt$, respectively. Within this framework, each of
\begin{equation}
\begin{gathered}
\intP \gamma \dvphiP \dv, \qquad \intdP \xis \dvphiP \da, \\[4pt]
\intdP \zeta \dvphidP \da, \qquad -\intdP \xis \dvphidP \da, \qquad \text{and} \qquad \intddP \tauds \dvphidP \ds,
\end{gathered}
\end{equation}
represents an external form of power expenditure, where $\gamma$ is the external bulk microforce, $\xis$ is the surface microtraction, $\zeta$ is the external surface microforce, and $\tauds$ is the edge microtraction. These power expenditures may be described as follows.
\begin{itemize}
  \item $\gamma \dvphiP$ represents the power expended on the atoms of $\prt$ by sources external to the body $\prt$; \\[4pt]
  \item $\xis \dvphiP$ represents the power expended across $\dprt$ by configurations neighboring the boundary of the body $\dprt$ and exterior to $\prt$; \\[4pt]
  \item $\zeta \dvphidP$ represents the power expended on the atoms of $\dprt$ by sources external to the boundary of the body $\dprt$ and not originated from $\prt$; \\[4pt]
  \item $-\xis \dvphidP$ represents the power expended on the atoms of $\dprt$ by sources external to the boundary of the body and originated from $\prt$; \\[4pt]
  \item $\tauds \dvphidP$ represents the power expended across $\ddprt$ by configurations neighboring the common boundary of the boundaries $\dprt^\pm$ of the body $\prt$ and exterior to both $\dprt$ and $\prt$.
\end{itemize}
Conversely, the internal power expenditure is given by the contribution of the following terms
\begin{equation}
\begin{gathered}
\intP \bs{\xi} \cdot \Grad \dvphiP \dv, \qquad - \intP \pi \dvphiP \dv, \\[4pt]
\intdP \bs{\tau} \cdot \Grads\dvphidP \da, \qquad \text{and} \qquad - \intdP \varpi \dvphidP \da,
\end{gathered}
\end{equation}
where $\bs{\xi}$ is the bulk microstress, $\pi$ is the internal bulk microforce, $\bs{\tau}$ is the surface microstress, and $\varpi$ is the internal surface microforce. We then base our treatment on the the virtual power principle formulation by Gurtin \cite{Gur02} and Fried \& Gurtin \cite{Fri06}. These works represent our point of departure to propose a generalized bulk-surface version of this principle. Through this suitable principle of virtual powers, we arrive at the microtractions presented in the external power, $\xis$ and $\tauds$, and the field equations.

Next, given the bulk and surface species fluxes, $\bsjP$ and $\bsjdP$, and the bulk and surface external rates of species production, $\sP$ and $\sdP$, we postulate the partwise species balances for $\prt$ and $\dprt$, where the balance on $\dprt$ is supplemented by a contribution originated from $\prt$ and given by
\begin{equation}
\intdP \beta \mskip3mu \bsjP \cdot \bs{n} \da.
\end{equation}
Then, with a suitable free-energy imbalance, we account for the rate at which energy is transferred to $\prt$ and $\dprt$ due to species transport to determine the constitutive relations and arrive at the following set of equations
\begin{equation}\label{eq:intro.governing}
\left\{
\begin{aligned}
\dvphiP &= \sP - \Div \bsjP, && \text{in } \prt, \\[4pt]
\muP &= - \Div\bs{\xi} - \gamma + \pvphiP \psiP, && \text{in } \prt, \\[4pt]
\dvphidP &= \beta \mskip3mu \bsjP \cdot \bs{n} + \sdP - \Divs \bsjdP - 2K \bsjdP \cdot \bs{n}, && \text{on } \dprt, \\[4pt]
\mudP &= - \Divs\bs{\tau} + 2K \bs{\tau} \cdot \bs{n} - \zeta + \bs{\xi} \cdot \bs{n} + \pvphidP \psidP, && \text{on } \dprt,
\end{aligned}
\right.
\end{equation}
and
\begin{equation}\label{eq:intro.constitutive}
\left\{
\begin{aligned}
\bs{\xi} &= \pgvphiP \psiP, && \text{in } \prt, \\[4pt]
\bsjP &= - \bsMP \mskip3mu \Grad\muP, && \text{in } \prt, \\[4pt]
\bs{\tau} &= \pgvphidP \psidP, && \text{on } \dprt, \\[4pt]
\bsjdP &= -\bsMdP \mskip3mu \Grads\mudP, && \text{on } \dprt,
\end{aligned}
\right.
\end{equation}
where $K \coloneqq - \fr{1}{2} \Divs \bs{n}$ is the mean curvature, $\pvphiP \coloneqq \partial/\partial\vphiP$, $\pvphidP \coloneqq \partial/\partial\vphidP$, $\pgvphiP \coloneqq \partial/\partial(\Grad\vphiP)$, and $\pgvphidP \coloneqq \partial/\partial(\Grads\vphidP)$. Additionally, $\bsMP$ and $\bsMdP$ are the bulk and surface mobility tensors, respectively.

Also, note that the appearance of bulk microstress $\bs{\xi}$ in the surface chemical potential $\mudP$, in Equation \eqref{eq:intro.governing}$_4$, results from the coupling at the principle of virtual powers' level, whereas the presence of the bulk species flux $\bsjP$ in the surface species time derivative $\dvphidP$, in Equation \eqref{eq:intro.governing}$_3$, results from the coupling at the surface species balance's level.

Aside from the fact that we present a new version of the principle of virtual powers and free-energy imbalance, there are three key differences between our continuum framework and previous works on dynamic boundary conditions. First, our theory is based on underlying mechanical principles. Second, our theory generalizes the resulting equations in \cite{Gol11,Kno21a}. Third, we consider that the boundary $\dprt$ may be endowed with a discontinuous normal field, allowing the assignment of edge microtractions. Lastly, in \cite{Kno21a}, a mixed (Robin) type of boundary condition for the chemical potentials is proposed for the model in \cite{Fis97,Gol11}. In addition to this boundary condition, we also include this type of mixed boundary condition for the microstructures, that is the phase fields.

\subsection{Differential tools}

In this subsection, we present helpful mathematical tools, from \cite{Esp21a,Esp21b}, to be used in the remainder of this study. We derive the relevant differential relations on a body $\bdy$ and a surface $\srf$.

Let $\prt$ be an arbitrary part embedded in a region $\bdy$ of a three-dimensional point space $\cl{E}$. With the coordinates $\tau^i$ $(i=1,2,3)$, the $i$th contravariant basis $\bs{g}^i$, and the conventional partial derivative $\partial_i\coloneqq\partial/\partial\tau^i$, let $\kappa$ and $\bs{\kappa}$ be, respectively, a smooth and a vector fields on $\bdy$. Then, within this setting, the gradient differential operator is defined as
\begin{equation}\label{eq:gradient.scalar}
\Grad\kappa \coloneqq \partial_i\kappa\mskip3mu\bs{g}^i.
\end{equation}

Next, consider a smooth surface $\srf\subset\prt$ oriented by the unit normal $\bs{n}$ at $\bs{x}\in\srf$. Let $\srf$ be parameterized by coordinates $\tau^p$ with $p=1,2$ and $\bs{z}$ be a smooth extension of $\srf$ along its normal $\bs{n}$ at $\bs{x}$ such that
\begin{equation}\label{eq:parameterization.surface.extension}
\bs{z}(\bs{x},\tau)\coloneqq\bs{x}+\tau\bs{n}(\bs{x}),\qquad\forall\,\bs{x}\in\srf,
\end{equation}
with $\tau$ representing the normal coordinate $n$ and taking values in an open interval of zero so that there exists a one-to-one mapping $\bs{z}\leftrightarrow(\bs{x},\tau)$. Such parameterization induces the local covariant basis
\begin{equation}\label{eq:covariant.basis}
\bs{g}_p\coloneqq\partial_p\bs{z}=\partial_p\bs{x}+\tau\partial_p\bs{n},\qquad\text{and}\qquad\bs{g}_n\coloneqq\partial_3\bs{z}=\bs{n}.
\end{equation}
With expression \eqref{eq:covariant.basis} at $\bs{x}=\bs{z}(\bs{x},0)$, we define
\begin{equation}
\bs{e}^p\coloneqq\bs{g}^p|_{\tau=0}.
\end{equation}
Thus,
\begin{equation}
\bs{e}_p=\partial_p\bs{x}.
\end{equation}
Furthermore, the contravariant $\bs{g}^p$ and covariant $\bs{g}_q$ bases satisfy
\begin{equation}\label{eq:covariant.contravariant.basis}
\bs{g}^p\cdot\bs{g}_q=\delta^{p}_{\cdot q}.
\end{equation}

Bearing in mind the parameterization \eqref{eq:parameterization.surface.extension}, consider the differential operators as follows. The gradient definition given in expression \eqref{eq:gradient.scalar} takes the form
\begin{equation}\label{eq:gradient.scalar.surface.extension}
\Grad\kappa=\partial_n\kappa\mskip3mu\bs{n}+\partial_p\kappa\mskip3mu\bs{e}^p.
\end{equation}
Next, let $\Prj{\bs{n}}\coloneqq\Prj{\bs{n}}(\bs{n})$ be the projector onto the plane defined by $\bs{n}$ at $\bs{x}\in\srf$ such that
\begin{equation}\label{eq:tan.projector}
\Prj{\bs{n}}\coloneqq\id-\bs{n}\otimes\bs{n}=\Prj{\bs{n}}^{\trans}.
\end{equation}
In view of the expression \eqref{eq:gradient.scalar.surface.extension} along with \eqref{eq:tan.projector}, the surface gradient is given by
\begin{equation}\label{eq:surface.gradient}
\Grads\kappa \coloneqq \partial_p\kappa\mskip3mu\bs{e}^p=\Prj{\bs{n}} \Grad\kappa,
\end{equation}
and the surface divergence by
\begin{equation}\label{eq:surface.divergence}
\Divs\bs{\kappa}=\partial_p\bs{\kappa}\cdot\bs{e}^p=\Grad\bs{\kappa}\twovdots\Prj{\bs{n}}.
\end{equation}
Then, the Laplace--Beltrami operator may be written as
\begin{equation}
\triangles\kappa \coloneqq \Divs\Grads\kappa = \Grad(\Prj{\bs{n}}\Grad\kappa)\twovdots\Prj{\bs{n}}.
\end{equation}

Lastly, for any smooth vector field $\bs{\kappa}$ on a smooth closed oriented surface $\srf$, the surface divergence theorem states that
\begin{equation}\label{eq:smooth.divs.theo.closed.S}
\int\limits_{\srf}\Divs(\Prj{\bs{n}} \bs{\kappa})\da = 0,
\end{equation}
whereas, owing to the lack of smoothness at an edge $\edg$, on a nonsmooth closed oriented surface $\srf$ with limiting outward unit tangent-normals $\bs{\nu}^+$ and $\bs{\nu}^-$ at $\edg$, the surface divergence theorem exhibits a \emph{surplus}, that is,
\begin{equation}\label{eq:nonsmooth.divs.theo.closed.S}
\int\limits_{\srf}\Divs(\Prj{\bs{n}} \bs{\kappa})\da=\int\limits_{\edg}\surp{\bs{\kappa}\cdot\bs{\nu}}\ds,
\end{equation}
where $\surp{\bs{\kappa}\cdot\bs{\nu}}\coloneqq\bs{\kappa}\cdot\bs{\nu}^++\bs{\kappa}\cdot\bs{\nu}^-$. Conversely, for open nonsmooth surfaces, we have to consider an extension of the surface divergence theorem \eqref{eq:nonsmooth.divs.theo.closed.S}, that is,
\begin{equation}\label{eq:nonsmooth.divs.theo.open.S}
\int\limits_{\srf}\Divs(\Prj{\bs{n}} \bs{\kappa})\da=\int\limits_{\dsrf}\bs{\kappa}\cdot\bs{\nu}\ds+\int\limits_{\edg}\surp{\bs{\kappa}\cdot\bs{\nu}}\ds.
\end{equation}

\section{Virtual power principle}
\label{sc:principle.virtual.powers}

We are now in a position to postulate the principle of virtual powers. Considering the power expenditures discussed in the previous section, the principle reads
\begin{equation}\label{eq:virtual.power.principle}
\cl{V}_{\mathrm{ext}}(\prt,\dprt;\chiP,\chidP) = \cl{V}_{\mathrm{int}}(\prt,\dprt;\chiP,\chidP),
\end{equation}
where $\chiP$ and $\chidP$ are two sufficiently smooth virtual fields defined, respectively, on $\prt$ and $\dprt$. The external and internal virtual power are, respectively, given by
\begin{equation}\label{eq:external.virtual.power}
\cl{V}_{\mathrm{ext}}(\prt,\dprt;\chiP,\chidP) = \intP \gamma \chiP \dv + \intdP (\zeta - \xis) \chidP \da + \intdP \xis \chiP \da + \intddP \tauds \chidP \ds,
\end{equation}
and
\begin{equation}\label{eq:internal.virtual.power}
\cl{V}_{\mathrm{int}}(\prt,\dprt;\chiP,\chidP) = \intP \bs{\xi} \cdot \Grad \chiP \dv - \intP \pi \chiP \dv + \intdP \bs{\tau} \cdot \Grads\chidP \da - \intdP \varpi \chidP \da.
\end{equation}
Next, we aim at deriving the explicit forms of the surface microtraction $\xis$ and the edge microtraction $\tauds$. Noting that $\bs{\tau} \cdot \Grads\chidP = \Prj{\bs{n}} \bs{\tau} \cdot \Grads\chidP$ while combining \eqref{eq:external.virtual.power} and \eqref{eq:internal.virtual.power} through \eqref{eq:virtual.power.principle} along with the divergence theorem and the surface divergence theorem for nonsmooth closed surfaces \eqref{eq:nonsmooth.divs.theo.closed.S}, we are led to
\begin{multline}
\intP \chiP (\Div\bs{\xi} + \pi + \gamma) \dv + \intdP \chiP (\xis - \bs{\xi} \cdot \bs{n}) \da \\[4pt]
+ \intdP \chidP (\Divs(\Prj{\bs{n}}\bs{\tau}) + \varpi + \zeta - \xis) \da + \intddP \chidP (\tauds - \surp{\bs{\tau} \cdot \bs{\nu}}) \ds = 0.
\end{multline}
Then, by variational arguments, the microtractions read
\begin{equation}\label{eq:microtractions}
\xis = \bs{\xi} \cdot \bs{n}, \qquad \text{and} \qquad \tauds = \surp{\bs{\tau} \cdot \bs{\nu}},
\end{equation}
while the bulk and surface field equations are given by
\begin{equation}\label{eq:field.equations}
\Div\bs{\xi} + \pi + \gamma = 0, \qquad \text{and} \qquad \Divs(\Prj{\bs{n}}\bs{\tau}) + \varpi + \zeta - \xis = 0.
\end{equation}
Note that the bulk microforce balance \eqref{eq:field.equations}$_1$ has the standard form proposed by Fried \& Gurtin \cite{Fri93}. However, the surface microforce balance \eqref{eq:field.equations}$_2$ has a contribution from the bulk, namely $\xis$. Additionally, the term $\Divs (\Prj{\bs{n}}\bs{\tau})$ may be split as $\Divs (\Prj{\bs{n}}\bs{\tau}) = \Divs \bs{\tau} + 2K \bs{\tau} \cdot \bs{n}$. Then, the surface microforce balance \eqref{eq:field.equations}$_2$ may be written as
\begin{equation}\label{eq:surface.field.equation}
\Divs\bs{\tau} + 2K \bs{\tau} \cdot \bs{n} + \varpi + \zeta - \xis = 0,
\end{equation}
for each smooth part of $\dprt$.

\section{Conserved species}
\label{sc:conserved.species}

We now account for the case where the bulk and surface phase fields, $\vphiP$ and $\vphidP$, represent the concentration of a conserved species. We therefore supplement the field equations \eqref{eq:field.equations} by two partwise species balances, that is, the bulk species balance
\begin{equation}\label{eq:bulk.balance.species}
\dot{\overline{\intP\vphiP\dv}} = \intP \sP \dv - \intdP \bsjP \cdot \bs{n} \da,
\end{equation}
and the surface species balance
\begin{equation}\label{eq:surface.balance.species}
\dot{\overline{\intdP\vphidP\da}} = \intdP \beta \mskip3mu \bsjP \cdot \bs{n} \da + \intdP \sdP \da - \intddP \surp{\bsjdP \cdot \bs{\nu}} \ds.
\end{equation}
The partiwise bulk and surface blance of species, respectively given by expressions \eqref{eq:bulk.balance.species} and \eqref{eq:surface.balance.species}, are motivated by the fact that we assume that the total balance of species satisfies
\begin{equation}
\dot{\overline{\intP\beta\mskip3mu\vphiP\dv}} + \dot{\overline{\intdP\vphidP\da}} = \intP \beta \mskip3mu \sP \dv + \intdP \sdP \da - \intddP \surp{\bsjdP \cdot \bs{\nu}} \ds.
\end{equation}

Using the divergence theorem and the surface divergence theorem for nonsmooth closed surfaces \eqref{eq:nonsmooth.divs.theo.closed.S} in expressions \eqref{eq:bulk.balance.species} and \eqref{eq:surface.balance.species}, respectively, followed by localization, we are led to
\begin{equation}\label{eq:pointwise.bulk.species.balance}
\dvphiP = \sP - \Div \bsjP,
\end{equation}
and
\begin{equation}\label{eq:pointwise.surface.species.balance}
\dvphidP = \beta \mskip3mu \bsjP \cdot \bs{n} + \sdP - \Divs(\Prj{\bs{n}}\bsjdP).
\end{equation}
Note that the bulk species balance has a standard form. However, the surface species balance has a contribution from the bulk, namely $\beta \mskip3mu \bsjP \cdot \bs{n}$. Additionally, the term $\Divs (\Prj{\bs{n}}\bsjdP)$ may be split as $\Divs (\Prj{\bs{n}}\bsjdP) = \Divs \bsjdP + 2K \bsjdP \cdot \bs{n}$. Then, the surface species balance \eqref{eq:pointwise.surface.species.balance} may be written as
\begin{equation}\label{eq:pointwise.surface.species.balance.alt}
\dvphidP = \beta \mskip3mu \bsjP \cdot \bs{n} + \sdP - \Divs\bsjdP - 2K \bsjdP \cdot \bs{n},
\end{equation}
for each smooth part of $\dprt$.

\section{Free-energy imbalance}
\label{sc:free.energy.imbalance}

First, note that the actual power is given by
\begin{equation}
\cl{W}_{\mathrm{ext}}(\prt,\dprt) \coloneqq \cl{V}_{\mathrm{ext}}(\prt,\dprt;\dvphiP,\dvphidP).
\end{equation}
In the free-energy imbalance, together with the external power expenditure, we account for the rate at which energy is transferred to $\prt$ and $\dprt$ due to species transport. Thus, the free-energy imbalance reads
\begin{align}\label{eq:partwise.free.energy.imbalance}
\dot{\overline{\intP\psiP\dv}} + \dot{\overline{\intdP\psidP\da}} \le {}& \cl{W}_{\mathrm{ext}}(\prt,\dprt) \nonumber\\[4pt]
&+ \intP \muP \sP \dv - \intdP \muP \bsjP \cdot \bs{n} \da \nonumber\\[4pt]
&+ \intdP \beta \mudP \bsjP \cdot \bs{n} \da + \intdP \mudP \sdP \da - \intddP \surp{\mudP \bsjdP \cdot \bs{\nu}} \ds.
\end{align}
Noting that $\Prj{\bs{n}}\bsjdP \cdot \Grads\mudP = \bsjdP \cdot \Grads\mudP$ and uncoupling $\dprt$ from $\prt$, for the sake of simplicity, we have that
\begin{equation}\label{eq:bulk.free.energy.imbalance.plain}
\dpsiP + (\pi - \muP) \dvphiP - \bs{\xi} \cdot \Grad\dvphiP + \bsjP \cdot \Grad\muP \le 0,
\end{equation}
and
\begin{equation}\label{eq:surface.free.energy.imbalance.plain}
\dpsidP + (\varpi - \mudP) \dvphidP - \bs{\tau} \cdot \Grads\dvphidP + \bsjdP \cdot \Grads\mudP \le 0.
\end{equation}
Additionally, assuming that the bulk and surface free-energy densities $\psiP$ and $\psidP$ are, respectively, given by constitutive response functions that are independent of $\muP$, $\mudP$, $\Grad\muP$, and $\Grads\mudP$
\begin{equation}
\psiP \coloneqq \psiP(\vphiP,\Grad\vphiP), \qquad \text{and} \qquad \psidP \coloneqq \psidP(\vphidP,\Grads\vphidP),
\end{equation}
we have that
\begin{equation}\label{eq:bulk.free.energy.total.derivative}
\dpsiP = \pvphiP \psiP \dvphiP + \pgvphiP \psiP (\Grad\vphiP)^{\bs{\dot{}}},
\end{equation}
and
\begin{equation}\label{eq:surface.free.energy.total.derivative}
\dpsidP = \pvphidP \psidP \dvphidP + \pgvphidP \psidP (\Grads\vphidP)^{\bs{\dot{}}}.
\end{equation}
Then, combining \eqref{eq:bulk.free.energy.imbalance.plain}, \eqref{eq:surface.free.energy.imbalance.plain}, \eqref{eq:bulk.free.energy.total.derivative}, and \eqref{eq:surface.free.energy.total.derivative}, we are led to two pointwise free-energy imbalances
\begin{equation}\label{eq:bulk.free.energy.imbalance}
(\muP - \pi - \pvphiP \psiP) \dvphiP + (\bs{\xi} - \pgvphiP \psiP) \cdot \Grad\dvphiP - \bsjP \cdot \Grad\muP \ge 0,
\end{equation}
and
\begin{equation}\label{eq:surface.free.energy.imbalance}
(\mudP - \varpi - \pvphidP \psidP) \dvphidP + (\bs{\tau} - \pgvphidP \psidP) \cdot \Grads\dvphidP - \bsjdP \cdot \Grads\mudP \ge 0.
\end{equation}
The bulk and surface free-energy imbalance equations, \eqref{eq:bulk.free.energy.imbalance} and \eqref{eq:surface.free.energy.imbalance}, serve to devise additional constitutive response functions in what follows.

\section{Additional constitutive response functions}
\label{sc:constitutive.relations}

We now assume that the set of independent variables is given by $\{\vphiP,\vphidP,\Grad\vphiP,\Grads\vphidP,\muP,\mudP\}$ while the set of dependent variables is $\{\pi,\varpi,\bs{\xi},\bs{\tau},\bsjP,\bsjdP\}$. Thus, we find that the local inequality \eqref{eq:bulk.free.energy.imbalance} and \eqref{eq:surface.free.energy.imbalance} are satisfied in all processes if and only if:
\begin{itemize}
\item The bulk and surface microstress $\bs{\xi}$ and $\bs{\tau}$ are, respectively, given by
\begin{equation}\label{eq:microstress}
\bs{\xi} \coloneqq \pgvphiP \psiP, \qquad \text{and} \qquad \bs{\tau} \coloneqq \pgvphidP \psidP.
\end{equation}
\item The internal bulk and surface microforces $\pi$ and $\varpi$ are, respectively, given by constitutive response functions that differs from the bulk and surface chemical potential by a contribution derived from the response functions $\psiP$ and $\psidP$
\begin{equation}\label{eq:internal.microforces}
\pi \coloneqq \muP - \pvphiP \psiP, \qquad \text{and} \qquad \varpi \coloneqq \mudP - \pvphidP \psidP.
\end{equation}
\item Granted that the bulk and surface species fluxes $\bsjP$ and $\bsjdP$ depend smoothly on the gradient of the bulk chemical potential, $\Grad\muP$, and the surface gradient of the surface chemical potentials, $\Grads\mudP$, these fluxes are, respectively, given by a constitutive response function of the form
\begin{equation}
\bsjP \coloneqq - \bsMP \Grad\muP, \qquad \text{and} \qquad \bsjdP \coloneqq -\bsMdP \Grads\mudP,
\end{equation}
where the mobility tensors $\bsMP$ and $\bsMdP$ must obey the residual dissipation inequalities
\begin{equation}
\Grad\muP \cdot \bsMP \Grad\muP \ge 0, \qquad \text{and} \qquad \Grads\mudP \cdot \bsMdP \Grads\mudP \ge 0.
\end{equation}
\end{itemize}
For the sake of simplicity, we let $\bsMP \coloneqq \mP \id$ and $\bsMdP \coloneqq \mdP \id$. With this choice for $\bsMdP$, the surface flux $\bsjdP$ remains proportional to $\Grads\mudP$ and therefore tangential to $\dprt$. Thus, the normal component of $\bsjdP$ appearing in expression \eqref{eq:pointwise.surface.species.balance.alt} vanishes. Also, note that nonlinear constitutive response functions for $\bsjP$ and $\bsjdP$ could be admissible as well, for instance, if $h \colon \bb{R}^3 \mapsto \bb{R}$ is a convex, differentiable function, one could assume $\bsjP \coloneqq \Grad h(\Grad \muP)$.

Important to what follows is the explicit form of the bulk and surface chemical potentials when using \eqref{eq:field.equations}$_1$ and \eqref{eq:surface.field.equation} in \eqref{eq:internal.microforces}. That is,
\begin{equation}
\muP = - \Div\bs{\xi} - \gamma + \pvphiP \psiP,
\end{equation}
and
\begin{equation}
\mudP = - \Divs\bs{\tau} - 2K \bs{\tau} \cdot \bs{n} - \zeta + \bs{\xi} \cdot \bs{n} + \pvphidP \psidP,
\end{equation}
which take the following form when considering \eqref{eq:microstress},
\begin{equation}\label{eq:bulk.chemical.potential}
\muP = \pvphiP \psiP - \Div\Big(\pgvphiP \psiP\Big) - \gamma,
\end{equation}
and
\begin{equation}\label{eq:surface.chemical.potential}
\mudP = \pvphidP \psidP- \Divs\Big(\pgvphidP \psidP\Big) - \zeta + \pgvphiP \psiP \cdot \bs{n},
\end{equation}
for each smooth part of $\dprt$. Note that the normal component of $\bs{\tau}$ vanishes for the free-energy function \eqref{eq:free.energy.functional} with \eqref{eq:microstress}$_2$.

In what follows, consider that $\prt \coloneqq \bdy$.

\subsection{Further connections: boundary conditions}

To be slightly more general, let us define different parts of the boundary $\dprt$. Let $\dprt^{\mathrm{dyn}}$ be the boundary with the dynamic bulk-surface interplay and $\dprt^{\mathrm{sta}}$ the static boundary such that $\dprt \coloneqq \dprt^{\mathrm{dyn}}\cup\dprt^{\mathrm{sta}}$ and $\dprt^{\mathrm{dyn}}\cap\dprt^{\mathrm{sta}}=\varnothing$. Also, let ${}_\circ\ddprt$ denote the boundary of the dynamic boundary $\dprt^{\mathrm{dyn}}$ while $\ddprt$ still denotes the edge along which the normal field is discontinuous. Thus, given the boundary conditions derived based upon thermodynamical principles in \cite{Esp20,Esp21a,Dud21}, we stipulate that the boundary conditions may be prescribed as follows.

The essential (Dirichlet) boundary conditions, that is, the assignment of microstructure, read
\begin{equation}\label{eq:dirichlet.phi.dP}
\vphiP (\bs{x},t) = \vphidP (\bs{x},t), \qquad \forall \bs{x} \in \dprt^{\mathrm{dyn}}_{\mathrm{ess}},
\end{equation}
and
\begin{equation}\label{eq:dirichlet.phi.ddP}
\vphidP (\bs{x},t) = \vphiddP^{\mathrm{env}} (\bs{x}), \qquad \forall \bs{x} \in \ddprt^{\mathrm{sta}}_{\mathrm{ess}} \cup {}_\circ\ddprt^{\mathrm{sta}}_{\mathrm{ess}},
\end{equation}
where the surface phase field $\vphidP$ is the action of the dynamic environment on $\dprt^{\mathrm{dyn}}_{\mathrm{ess}}$ and $\vphiddP^{\mathrm{env}}$ is the action of the static environment on $\ddprt^{\mathrm{sta}}_{\mathrm{ess}}$. On a static environment, expression \eqref{eq:dirichlet.phi.dP} takes the form
\begin{equation}\label{eq:dirichlet.phi.dP.sta}
\vphiP (\bs{x},t) = \vphidP^{\mathrm{env}} (\bs{x}), \qquad \forall \bs{x} \in \dprt^{\mathrm{sta}}_{\mathrm{ess}},
\end{equation}
where $\vphidP^{\mathrm{env}}$ is the action of the static environment on $\dprt^{\mathrm{sta}}_{\mathrm{ess}}$.

Instead, we may opt for the natural (Neumann) boundary conditions, that is, the assignment of microtractions. Then, we have
\begin{equation}
\xis (\bs{x},t) = \xis^{\mathrm{env}} (\bs{x}), \qquad \forall \bs{x} \in \dprt^{\mathrm{sta}}_{\mathrm{nat}},
\end{equation}
and
\begin{equation}\label{eq:edge.boundary.condition.surface.flux.junction}
\tauds (\bs{x},t) = \tauds^{\mathrm{env}}(\bs{x}), \qquad \forall \bs{x} \in \ddprt^{\mathrm{sta}}_{\mathrm{nat}},
\end{equation}
and
\begin{equation}\label{eq:edge.boundary.condition.surface.flux.boundary}
\otauds (\bs{x},t) = \tauds^{\mathrm{env}}(\bs{x}), \qquad \forall \bs{x} \in {}_\circ\ddprt^{\mathrm{sta}}_{\mathrm{ess}},
\end{equation}
where $\xis = \bs{\xi} \cdot \bs{n}$, $\tauds = \surp{\bs{\tau} \cdot \bs{\nu}}$ on $\ddprt^{\mathrm{sta}}_{\mathrm{nat}} $, $\otauds = \bs{\tau} \cdot \bs{\nu}$ on ${}_\circ\ddprt^{\mathrm{sta}}_{\mathrm{nat}}$ due to \eqref{eq:nonsmooth.divs.theo.open.S}, and $\xis^{\mathrm{env}}$ and $\tauds^{\mathrm{env}}$ are the actions of the static environment, respectively, on $\dprt^{\mathrm{sta}}_{\mathrm{nat}}$ and $\ddprt^{\mathrm{sta}}_{\mathrm{nat}} \cup {}_\circ\ddprt^{\mathrm{sta}}_{\mathrm{nat}}$.

As for the bulk species balance, as an essential boundary condition, we may prescribe
\begin{equation}\label{eq:dirichlet.mu.dP}
\muP (\bs{x},t) = \beta \mudP(\bs{x},t), \qquad \forall \bs{x} \in \dprt^{\mathrm{dyn}}_{\mathrm{ess}},
\end{equation}
where $\beta \mudP$ is the action of the dynamic environment on $\dprt^{\mathrm{sta}}_{\mathrm{nat}}$. Expression \eqref{eq:dirichlet.mu.dP}, on a static environment, takes the form
\begin{equation}\label{eq:dirichlet.mu.dP.sta}
\muP (\bs{x},t) = \mudP^{\mathrm{env}}(\bs{x}), \qquad \forall \bs{x} \in \dprt^{\mathrm{sta}}_{\mathrm{ess}},
\end{equation}
where $\mudP^{\mathrm{env}}$ is the action of the static environment on $\dprt^{\mathrm{sta}}_{\mathrm{ess}}$.

Instead, as a natural boundary condition, we may opt for
\begin{equation}
\bsjP (\bs{x},t) \cdot \bs{n} = - \jmath_{\scriptscriptstyle\dprt}^{\mathrm{env}} (\bs{x}), \qquad \forall \bs{x} \in \dprt^{\mathrm{sta}}_{\mathrm{nat}},
\end{equation}
where $\jmath_{\scriptscriptstyle\dprt}^{\mathrm{env}}$ is the action of the static environment on $\dprt^{\mathrm{sta}}_{\mathrm{nat}}$ through the normal component of the flux $\bsjP$. Conversely, for the surface species balance, as an essential boundary condition, we may prescribe
\begin{equation}
\mudP (\bs{x},t) = \muddP^{\mathrm{env}}(\bs{x}), \qquad \forall \bs{x} \in \ddprt^{\mathrm{sta}}_{\mathrm{ess}} \cup {}_\circ\ddprt^{\mathrm{sta}}_{\mathrm{ess}},
\end{equation}
where $\muddP^{\mathrm{env}}$ is the action of the static environment on $\ddprt^{\mathrm{sta}}_{\mathrm{ess}} \cup {}_\circ\ddprt^{\mathrm{sta}}_{\mathrm{ess}}$, whereas, as a natural boundary condition, we may opt for
\begin{equation}\label{eq:edge.boundary.condition.bulk.flux}
\surp{\bsjdP (\bs{x},t) \cdot \bs{\nu}} = - \jmath_{\scriptscriptstyle\ddprt}^{\mathrm{env}} (\bs{x}), \qquad \forall \bs{x} \in \ddprt^{\mathrm{sta}}_{\mathrm{nat}},
\end{equation}
and, due to \eqref{eq:nonsmooth.divs.theo.open.S},
\begin{equation}
\bsjdP (\bs{x},t) \cdot \bs{\nu} = - \jmath_{\scriptscriptstyle\ddprt}^{\mathrm{env}} (\bs{x}), \qquad \forall \bs{x} \in {}_\circ\ddprt^{\mathrm{sta}}_{\mathrm{nat}},
\end{equation}
where $\jmath_{\scriptscriptstyle\ddprt}^{\mathrm{env}}$ is the action of the static environment on $\ddprt^{\mathrm{sta}}_{\mathrm{nat}}$ through the tangent-normal component of the flux $\bsjdP$.

For mixed boundary conditions on static environments, the reader is referred to \cite{Esp21a,Esp21b}. Here, we restrict attention to the mixed boundary conditions on $\dprt^{\mathrm{dyn}}$. Specifically, we invoke relation proposed by Fried \& Gurtin \cite[surface free-energy imbalance (92)]{Fri06} and stipulate that
\begin{equation}\label{eq:free.energy.environment}
\cl{T}_{\mathrm{surf}}(-\dprt) + \cl{T}_{\mathrm{env}}(\dprt) \ge 0,
\end{equation}
where $\cl{T}_{\mathrm{surf}}(-\dprt)$ combines the power expended on $\dprt$ by the material inside $\prt$ and the rate at which energy is transferred to $\prt$ and $\dprt$, whereas $\cl{T}_{\mathrm{env}}(\dprt)$ combines the power expended by the environment on $\dprt$ and the rate at which energy is transferred from the environment to $\dprt$. For $\dprt^{\mathrm{dyn}}$, we here define
\begin{align}\label{eq:power.surface.dyn}
\cl{T}_{\mathrm{surf}}(-\dprt^{\mathrm{dyn}}) \coloneqq {}& - \intdP (\dvphiP - \dvphidP) \xis \da - \intddP \tauds \dvphidP \ds + \intoddP \otauds \dvphidP \ds \nonumber\\[4pt]
& - \intdP (\beta \mudP - \muP) \bsjP \cdot \bs{n} \da + \intddP \surp{\mudP \bsjdP \cdot \bs{\nu}} \ds \nonumber\\[4pt]
& + \intoddP \mudP \bsjdP \cdot \bs{\nu} \ds,
\end{align}
where, owing to \eqref{eq:nonsmooth.divs.theo.open.S}, $\otauds \coloneqq \bs{\tau} \cdot \bs{\nu}$ is the analogous of $\tauds$ but developed on ${}_\circ\ddprt$. We also define
\begin{align}\label{eq:power.env.dyn}
\cl{T}_{\mathrm{env}}(\dprt^{\mathrm{dyn}}) \coloneqq {}& \intddP \tauds^{\mathrm{env}} \dvphiddP^{\mathrm{env}} \ds + \intoddP \otauds^{\mathrm{env}} \dvphiddP^{\mathrm{env}} \ds \nonumber\\[4pt]
& - \intddP \muddP^{\mathrm{env}} \jmath_{\scriptscriptstyle\ddprt}^{\mathrm{env}} \ds - \intoddP \muddP^{\mathrm{env}} \jmath_{\scriptscriptstyle\ddprt}^{\mathrm{env}} \ds.
\end{align}

Now, we consider that on $\ddprt \cup {}_\circ\ddprt$, $\tauds^{\mathrm{env}} = \tauds$, $\dvphiddP^{\mathrm{env}} = \dvphidP$, $\muddP^{\mathrm{env}} = \mudP$, whereas on $\ddprt$, $\jmath_{\scriptscriptstyle\ddprt}^{\mathrm{env}} = \surp{\bsjdP \cdot \bs{\nu}}$ and on ${}_\circ\ddprt$, $\jmath_{\scriptscriptstyle\ddprt}^{\mathrm{env}} = \bsjdP \cdot \bs{\nu}$. Thus, expression \eqref{eq:free.energy.environment} reads
\begin{equation}\label{eq:free.energy.environment.mix}
- \intdP \big( (\dvphiP - \dvphidP) \mskip3mu \xis + (\beta \mudP - \muP) \mskip3mu \bsjP \cdot \bs{n} \big) \da \ge 0.
\end{equation}
Unclupling this expression, we have that
\begin{equation}\label{eq:mix.boundary}
\intdP (\dvphiP - \dvphidP) \mskip3mu \xis \da \le 0, \qquad \text{and} \qquad \intdP (\beta \mudP - \muP) \mskip3mu \bsjP \cdot \bs{n} \da \le 0\footnotemark.
\end{equation}
\footnotetext{This type of condition was derived by Espath \& Calo \cite[Equation (157)]{Esp21a} based upon mechanical and thermodynamical arguments.} Note that, the terms in \eqref{eq:mix.boundary} are dissipative. That is, as a mixed boundary condition, expressions \eqref{eq:mix.boundary} read
\begin{equation}\label{eq:mix.bulk.phase}
\bs{\xi} (\bs{x},t) \cdot \bs{n} = \fr{1}{L_\varphi} (\dvphidP - \dvphiP), \qquad \forall \bs{x} \in \dprt^{\mathrm{dyn}}_{\mathrm{mix}},
\end{equation}
and
\begin{equation}\label{eq:mix.bulk.flux}
\bsjP (\bs{x},t) \cdot \bs{n} = - \fr{1}{L_\mu} (\beta \mudP - \muP), \qquad \forall \bs{x} \in \dprt^{\mathrm{dyn}}_{\mathrm{mix}},
\end{equation}
where $L_\varphi, L_\mu > 0$.

\subsection{Specialized equations}

In view of the free-energy functional \eqref{eq:free.energy.functional}, and expressions \eqref{eq:pointwise.bulk.species.balance}, \eqref{eq:pointwise.surface.species.balance.alt}, \eqref{eq:bulk.chemical.potential}, and \eqref{eq:surface.chemical.potential}, our theory renders the following set of equations
\begin{equation}\label{eq:recovering.original}
\left\{
\begin{aligned}
\dvphiP &= \sP + \mP\triangle\muP, && \text{in } \prt \\[4pt]
\muP &= -\epsilon \triangle\vphiP + \fr{1}{\epsilon} f^\prime (\vphiP) - \gamma, && \text{in } \prt, \\[4pt]
\dvphidP &= \sdP + \mdP\triangles\mudP - \beta \mP \mskip3mu \partial_n\muP, && \text{on } \dprt, \\[4pt]
\mudP &= - \iota \delta \triangles\vphidP + \fr{1}{\delta} g^\prime (\vphidP) + \epsilon \mskip3mu \partial_n \vphiP - \zeta && \text{on } \dprt.
\end{aligned}
\right.
\end{equation}

The boundary conditions, may be summarized as follows:
\begin{equation}
\forall \bs{x} \in \dprt^{\mathrm{dyn}}
\left\{
\begin{aligned}
\vphiP &= \vphidP, \qquad \text{or} \qquad \epsilon \mskip3mu \partial_n \vphiP = \fr{1}{L_\varphi} (\dvphidP - \dvphiP), \\[4pt]
\muP &= \beta \mudP, \qquad \text{or} \qquad - \mP \mskip3mu \partial_n \muP = - \fr{1}{L_\mu} (\beta \mudP - \muP),
\end{aligned}
\right.
\end{equation}
and
\begin{equation}
\forall \bs{x} \in \dprt^{\mathrm{sta}}
\left\{
\begin{aligned}
\vphiP &= \vphidP^{\mathrm{env}}, \qquad \text{or} \qquad \epsilon \mskip3mu \partial_n \vphiP = \xis^{\mathrm{env}}, \\[4pt]
\muP &= \mudP^{\mathrm{env}}, \qquad \text{or} \qquad - \mP \mskip3mu \partial_n \muP = - \jmath_{\scriptscriptstyle\dprt}^{\mathrm{env}},
\end{aligned}
\right.
\end{equation}
and
\begin{equation}
\forall \bs{x} \in \ddprt^{\mathrm{sta}}
\left\{
\begin{aligned}
\vphidP &= \vphiddP^{\mathrm{env}}, \qquad \text{or} \qquad \iota \delta \mskip3mu \surp{\partial_{\nu} \vphidP} = \tauds^{\mathrm{env}}, \\[4pt]
\mudP &= \muddP^{\mathrm{env}}, \qquad \text{or} \qquad - \mdP \surp{\partial_\nu \mudP} = - \jmath_{\scriptscriptstyle\ddprt}^{\mathrm{env}},
\end{aligned}
\right.
\end{equation}
and
\begin{equation}
\forall \bs{x} \in {}_\circ\ddprt^{\mathrm{sta}}
\left\{
\begin{aligned}
\vphidP &= \vphiddP^{\mathrm{env}}, \qquad \text{or} \qquad \iota \delta \mskip3mu \partial_{\nu} \vphidP = \tauds^{\mathrm{env}}, \\[4pt]
\mudP &= \muddP^{\mathrm{env}}, \qquad \text{or} \qquad - \mdP \mskip3mu \partial_\nu \mudP = - \jmath_{\scriptscriptstyle\ddprt}^{\mathrm{env}}.
\end{aligned}
\right.
\end{equation}

\section{Decay relations}
\label{sc:lyapunov}

We now aim to establish Lyapunov-decay relations for the case where all the boundary $\dprt$ is dynamic and of the mixed type for both the microstructure and chemical potential. Thus, in view of \eqref{eq:bulk.free.energy.total.derivative} and \eqref{eq:surface.free.energy.total.derivative} combined with the constitutive relations for the bulk and surface microstresses \eqref{eq:microstress}, we have that
\begin{align}
\intP \dpsiP \dv + \intdP \dpsidP \da ={}& \intP \big( \pvphiP \psiP \dvphiP + \pgvphiP \psiP (\Grad\vphiP)^{\bs{\dot{}}} \mskip2mu \big) \dv \nonumber \\[4pt]
&+ \intdP \big( \pvphidP \psidP \dvphidP + \pgvphidP \psidP (\Grads\vphidP)^{\bs{\dot{}}} \mskip2mu \big) \da, \nonumber \\[4pt]
={}& \intP \big( \pvphiP \psiP \dvphiP + \bs{\xi} \cdot \Grad\dvphiP \big) \dv \nonumber \\[4pt]
&+ \intdP \big( \pvphidP \psidP \dvphidP + \Prj{\bs{n}}\bs{\tau} \cdot \Grads\dvphidP \big) \da, \nonumber \\[4pt]
={}& \intP \big( (\pvphiP \psiP - \Div \bs{\xi}) \dvphiP + \Div (\dvphiP \mskip3mu \bs{\xi}) \big) \dv \nonumber \\[4pt]
&+ \intdP \big( (\pvphidP \psidP - \Divs (\Prj{\bs{n}}\bs{\tau})) \dvphidP + \Divs(\dvphidP \mskip3mu \Prj{\bs{n}}\bs{\tau}) \big) \da.
\end{align}
Next, using the pointwise balances of microforces \eqref{eq:field.equations}, we arrive at
\begin{align}\label{eq:free.energy.decay}
\intP \dpsiP \dv + \intdP \dpsidP \da ={}& \intP \big( (\pvphiP \psiP + \pi + \gamma) \dvphiP + \Div (\dvphiP \mskip3mu \bs{\xi}) \big) \dv \nonumber \\[4pt]
& + \intdP \big( (\pvphidP \psidP + \varpi + \zeta - \xis) \dvphidP + \Divs(\dvphidP \mskip3mu \Prj{\bs{n}}\bs{\tau}) \big) \da.
\end{align}
Note that relation \eqref{eq:free.energy.decay} holds whether or not the species transports in the bulk and on the surface are present or not. Conversely, in view of expression \eqref{eq:microtractions}$_1$, the surface microtraction, and accounting for the bulk and surface internal microforces, given by expressions \eqref{eq:internal.microforces}, in \eqref{eq:free.energy.decay}, we are led to
\begin{align}\label{eq:free.energy.decay.chemical}
\intP \dpsiP \dv + \intdP \dpsidP \da ={}& \intP \big( (\muP + \gamma) \dvphiP + \Div (\dvphiP \mskip3mu \bs{\xi}) \big) \dv \nonumber \\[4pt]
& + \intdP \big( (\mudP + \zeta - \bs{\xi} \cdot \bs{n}) \dvphidP + \Divs(\dvphidP \mskip3mu \Prj{\bs{n}}\bs{\tau}) \big) \da.
\end{align}
We now use the bulk and surface species balances, respectively given by \eqref{eq:pointwise.bulk.species.balance} and \eqref{eq:pointwise.surface.species.balance} in expression \eqref{eq:free.energy.decay.chemical}, to arrive at
\begin{align}
\intP \dpsiP \dv + \intdP \dpsidP \da ={}& \intP \big( - \Grad \muP \cdot \bsMP \Grad \muP + \muP \sP + \gamma \dvphiP + \Div (\dvphiP \mskip3mu \bs{\xi} - \muP \bsjP) \big) \dv \nonumber \\[4pt]
& + \intdP \big( - \Grads \mudP \cdot \bsMdP \Grads \mudP + \mudP (\beta \mskip3mu \bsjP \cdot \bs{n} + \sdP) \big) \da \nonumber \\[4pt]
& + \intdP \big( (\zeta - \bs{\xi} \cdot \bs{n}) \dvphidP + \Divs(\dvphidP \mskip3mu \Prj{\bs{n}}\bs{\tau} - \mudP \mskip3mu \Prj{\bs{n}}\bsjdP) \big) \da.
\end{align}
Then, using the divergence theorem and the surface divergence theorem for closed nonsmooth surfaces \eqref{eq:nonsmooth.divs.theo.closed.S}, we are led to
\begin{align}\label{eq:free.energy.decay.transport}
\intP \dpsiP \dv + \intdP \dpsidP \da ={}& \intP \big( - \Grad \muP \cdot \bsMP \Grad \muP + \muP \sP + \gamma \dvphiP \big) \dv \nonumber \\[4pt]
& + \intdP \big( - \Grads \mudP \cdot \bsMdP \Grads \mudP + \mudP\sdP + (\beta \mudP - \muP) \mskip3mu \bsjP \cdot \bs{n} \big) \da \nonumber \\[4pt]
& + \intdP \big( \zeta \dvphidP + (\dvphiP - \dvphidP) \mskip3mu \bs{\xi} \cdot \bs{n} \big) \da \nonumber \\[4pt]
& + \intddP \big( \surp{\dvphidP \mskip3mu \bs{\tau} \cdot \bs{\nu}} - \surp{\mudP \mskip3mu \bsjdP \cdot \bs{\nu}} \big) \ds.
\end{align}

\subsection{Decay relations for mixed boundary conditions}

For a mixed boundary condition on $\dprt \coloneqq \dprt^{\mathrm{dyn}}_{\mathrm{mix}}$ given by expressions \eqref{eq:mix.bulk.phase} and \eqref{eq:mix.bulk.flux} along with the boundary-edge conditions \eqref{eq:edge.boundary.condition.surface.flux.junction} and \eqref{eq:edge.boundary.condition.bulk.flux}, we obtained the final relation
\begin{align}
\intP \dpsiP \dv + \intdP \dpsidP \da ={}& \intP \big( - \Grad \muP \cdot \bsMP \Grad \muP + \muP \sP + \gamma \dvphiP \big) \dv \nonumber \\[4pt]
& + \intdP \big( - \Grads \mudP \cdot \bsMdP \Grads \mudP + \mudP\sdP + \zeta \dvphidP \big) \da \nonumber \\[4pt]
& + \intdP \big( - \fr{1}{L_\varphi} (\dvphidP - \dvphiP)^2 - \fr{1}{L_\mu} (\beta \mudP - \muP)^2 \big) \da + \intddP \big( \dvphidP \mskip3mu \tauds^{\mathrm{env}} + \mudP \mskip3mu \jmath_{\scriptscriptstyle\ddprt}^{\mathrm{env}} \big) \ds.
\end{align}
Thus, the final Lyapunov-decay relation is
\begin{align}\label{eq:free.energy.decay.transport.mix}
\intP \dpsiP \dv + \intdP \dpsidP \da \le{}& \intP \big( \muP \sP + \gamma \dvphiP \big) \dv + \intdP \big( \mudP\sdP + \zeta \dvphidP \big) \da \nonumber \\[4pt]
& + \intddP \big( \dvphidP \mskip3mu \tauds^{\mathrm{env}} + \mudP \mskip3mu \jmath_{\scriptscriptstyle\ddprt}^{\mathrm{env}} \big) \ds.
\end{align}

\section{Acknowledgements}

The author would like to thank Gabriel Nogueira de Castro for rendering the image in Figure \ref{fg:geometry}. This work was partially supported by the Flexible Interdisciplinary Research Collaboration Fund at the University of Nottingham Project ID 7466664.

\section{Declarations: Funding and/or Conflicts of interests/Competing interests \& Data availability}

The authors have no conflicts to disclose.

Data sharing is not applicable to this article as no new data were created or analyzed in this study.


%


\footnotesize


\end{document}